\documentclass{article}
\usepackage{epsfig}

\newcommand{\sect}[1]{\S\ref{sect.#1}}
\newcommand{\eq}[1]{Eq.~\ref{eq.#1}}
\newcommand{\fig}[1]{Fig.~\ref{#1}}

\newcommand{\sectlabel}[1]{\label{sect.#1}}
\newcommand{\eqlabel}[1]{\label{eq.#1}}

\newcommand{\expect}[1]{ {\left\langle #1 \right\rangle} }

\newcommand{\Psoln}{ P_{\rm soln} }
\newcommand{\Psoluble}{ { P_{\rm soluble} }}
\newcommand{\Nclauses}{ N_{\rm clauses} }  % used as \Nclauses^m so no extra braces

\newcommand{\states}[2]{ {v_{#1}(#2)} }

\newcommand{\AtMost}[1] { O \left( #1 \right) }
\newcommand{\AtLeast}[1] { \Omega \left( #1 \right) }
\newcommand{\Same}[1] { \Theta \left( #1 \right) }

\begin{document}
\title{Solving Random Satisfiability Problems with Quantum Computers}
\author{Tad Hogg\\
Xerox Palo Alto Research Center\\
Palo Alto, CA 94304\\
hogg@parc.xerox.com}

\maketitle

\begin{abstract}
Quantum computer algorithms can exploit the structure of random
satisfiability problems. This paper extends a previous empirical
evaluation of such an algorithm and gives an approximate asymptotic
analysis accounting for both the average and variation of amplitudes
among search states with the same costs. The analysis predicts good
performance, on average, for a variety of problems including those
near a phase transition associated with a high
concentration of hard cases. Based on empirical evaluation for small
problems, modifying the algorithm in light of
this analysis improves its performance. The algorithm improves on both GSAT, a commonly used
conventional heuristic, and quantum algorithms ignoring problem
structure.
\end{abstract}

% for revtex style
%\pacs{03.67.Lx, 89.70.+c, 02.70.-c}

%%%%%%%%%%%%%%%%%%%%%%%%%%%%%%%%%%%%%%%%%%%%%%%%%%%%%%%%%%%%%%%%
\section{Introduction}

Peter Shor's polynomial-time factoring
algorithm~\cite{shor94,chuang95} showed quantum
computers~\cite{deutsch85,divincenzo95,feynman96,steane98} efficiently
solve an important problem thought to require exponential time on our
current, ``classical'', machines. Can quantum computers significantly
improve other apparently intractable problems? At first sight,
combinatorial searches, such as arise in scheduling, theorem proving,
cryptography, genetics and statistical physics, are one
possibility. This is because many such searches are ``nondeterministic
polynomial'' (NP) problems~\cite{garey79}, which have a rapid test of
whether a candidate solution is in fact a solution and an exponential
growth in the number of candidates with the size of the
problem. Quantum computers can test all candidates in superposition
with about as many operations as a classical machine uses to test just
one, suggesting large improvements are possible. Unfortunately, the
difficulty of extracting a solution from the superposition appears to
preclude rapid solution of of at least some NP
problems~\cite{bennett94}.

Nevertheless, quantum computers may offer substantial improvement for
{\em typical} searches encountered in practice. For instance,
constraint satisfaction problems~\cite{mackworth92} consist of
constraints on the values various combinations of variables can
take. A candidate solution for such problems can not only be evaluated
in terms of {\em whether} it satisfies all the constraints, but also
in terms of {\em how many} constraints it violates. This additional
information is often a useful guide to finding solutions, providing
the basis for conventional heuristic searches.  Such heuristics
are substantially better
than simpler techniques ignoring problem structure.  For heuristics
consisting of repeated independent trials, Grover's amplitude
amplification~\cite{grover96} gives a quadratic speedup with quantum
computers~\cite{brassard98}, the best possible improvement for quantum methods
based only on the test of whether candidates are
solutions~\cite{bennett94}.

Any possibility of greater speedup requires a quantum algorithm using
additional problem properties.  For some small or relatively easy
problems such algorithms perform well~\cite{hogg98,spector99}.  More
generally, quantum methods readily exploit precise information on
states' distances to a solution~\cite{grover97b}, but such information
is not readily available for hard searches. Thus an important question
is whether, and to what extent, quantum computers can exploit readily
computed properties of hard search problems. In particular, can they
perform significantly more efficiently than classical heuristic
methods?

This paper discusses a previous structured quantum search
algorithm~\cite{hogg00}, based on evaluating, in superposition, the
number of conflicts in all search states. The paper extends empirical
evaluation of the algorithm's behavior for a class of hard search
problems and compares it to a version of amplitude amplification not
requiring prior knowledge of the number of
solutions~\cite{boyer96}. Furthermore, the paper gives an approximate
asymptotic performance analysis that includes variation among
amplitudes associated with states with the same number of
conflicts. Specifically, the next two sections describe a class of
hard search problems and the form of the quantum algorithm. The
remainder of the paper presents the extended asymptotic analysis and
compares with actual behavior based on small problem sizes feasible to
evaluate via simulation on conventional machines.

As a note on notation, to compare the growth rates of various
functions we use~\cite{graham94} $f = \AtMost{g}$ to indicate that $f$
grows no faster than $g$ as a function of $n$ when $n \rightarrow
\infty$. Conversely, $f = \AtLeast{g}$ means $f$ grows at least as
fast as $g$, and $f = \Same{g}$ means both functions grow at the same
rate.

\section{An Ensemble of Hard Satisfiability Problems}

Heuristics are often too complicated to allow exact analytical
evaluation of their performance. Instead, they are usually evaluated
empirically on a sample of problems. Such a test requires a
hard problem ensemble, i.e., a class of problem instances and
associated probability distribution for their selection with a high
concentration of hard cases. For practical use, instances of the
ensemble should be computationally easy to generate. Since typical
instances of NP problems are often much easier than worst case
analyses suggest, defining such ensembles is not trivial.
Fortunately, such ensembles exist for a variety of NP-complete search
problems~\cite{cheeseman91,kirkpatrick94,hogg96d}.  Significantly,
problems from such ensembles, associated with abrupt ``phase
transitions'' in behavior, are particularly difficult for a variety of
heuristics, on average. They thus provide good test cases.

The $k$-satisfiability ($k$-SAT) problem provides one example.  It
consists of $n$ Boolean variables and $m$ clauses. A clause is a
logical OR of $k$ variables, each of which may be negated. A solution
is an assignment, i.e., a value, true or false, for each variable,
satisfying all the clauses. An assignment is said to conflict with any
clause it doesn't satisfy.  An example 2-SAT problem instance with 3
variables and 2 clauses is $v_1$ OR (NOT $v_2$) and $v_2$ OR $v_3$,
which has 4 solutions, e.g., $v_1={\rm false}$, $v_2={\rm false}$ and
$v_3={\rm true}$.  For $k\ge 3$, $k$-SAT is
NP-complete~\cite{garey79}, i.e., is among the most difficult NP
problems.

For assignments $r$ and $s$, which can be viewed as bit-vectors of
length $n$, let $d(r,s)$ be the Hamming distance between them, i.e.,
the number of variables they assign different values. Let $c(s)$
denote the number of the $m$ clauses conflicting with $s$, which
depends on the particular problem instance considered. The quantity
$c(s)$ can also be thought of as the cost associated with the
assignment, and those with zero cost are solutions.

The random $k$-SAT ensemble with given $n$ and $m$ consists of
instances whose $m$ clauses are selected uniformly at
random. Specifically, for each clause, a set of $k$ variables is
selected randomly from among the $n
\choose k$ possibilities. Then each of the selected variables is
negated with probability $1/2$ to produce the clause. Thus each of the
$m$ clauses is selected, with replacement, uniformly from among the
$\Nclauses = {n \choose k} 2^k$ possible clauses. The difficulty of
solving such randomly generated problems varies greatly from one
instance to the next. This ensemble has a high concentration of hard
instances when $\mu \equiv m/n$ is near a phase transition in search
difficulty~\cite{cheeseman91,kirkpatrick94,hogg96d}. At this
transition, the fraction of soluble instances drops abruptly from near
1 to near 0. For random 3-SAT this transition is at about $\mu=4.25$,
the value used for the results presented here as well as extensive
prior studies of classical heuristics for SAT.
For soluble problems near the transition, the number of solutions $S$ is
exponentially large but a tiny fraction of all states, i.e., $S/2^n$ is
exponentially small.

\section{The Algorithm}

The quantum algorithm examined here~\cite{hogg00} has the same general
form as amplitude amplification~\cite{grover96} but with amplitude
phase adjustments based on the state costs and the problem ensemble
parameters, i.e., $n$, $k$ and $m$ for random $k$-SAT.  Importantly,
the algorithm does not require characteristics of the problem instance
that are costly to compute, e.g., the number of solutions.

The overall algorithm consists of a series of trials, each operating
with superpositions of all $2^n$ assignments. Superpositions
correspond to vectors with an amplitude for each assignment. After
each trial, a measurement produces a single assignment. Trials repeat
until a solution is found. Quantum coherence need persist only for the
duration of each trial, rather than over all trials. For the case
considered here, this duration grows linearly with $n$ thereby placing
less stringent coherence requirements on the hardware than amplitude
amplification whose trial duration grows exponentially with $n$ for
hard problems (because, for hard problems, the number of solutions is
an exponentially small fraction of the total number of states).

A trial performs a series of $j$ steps. Each step evaluates the costs
associated with all assignments and mixes amplitudes among them based
on their Hamming distances.  Starting with an equal superposition of
all $2^n$ assignments, i.e., $\psi^{(0)}_s = 2^{-n/2}$, the
superposition vector $\psi^{(j)}$ after $j$ steps is
\begin{equation}\eqlabel{map}
\psi^{(j)} = U^{(j)} P^{(j)} \ldots U^{(1)} P^{(1)} \psi^{(0)}
\end{equation}
The algorithm involves two types of matrices: the diagonal phase adjustments 
$P^{(h)}$, depending on the particular problem instance, and the matrix $U^{(h)}$, mixing amplitudes among
states without regard to the particular instance.

Specifically, $P^{(h)}$ is diagonal with $P^{(h)}_{ss} = e^{i \pi
\rho(h,c(s))}$ where $c(s)$ is the number of conflicts in assignment
$s$ and $\rho$ is an arbitrary computationally-efficient real-valued
function. Since $c(s)$ itself is efficiently evaluated (by comparing
the state with each of the $m$ clauses) and has only $m+1=\Same{n}$
possible values, $0,\ldots,m$, quantum computers efficiently implement
this matrix operation~\cite{hogg98b} as a generalization of the
technique used for amplitude amplification.

Viewing assignments as strings of $n$ bits, let $W$ be the
Walsh-transform, $W_{rs} = 2^{-n/2} (-1)^{|r \wedge s|}$ where $|r
\wedge s|$ is the number of 1's the two assignments have in common. We
define the mixing matrix as $U^{(h)} = W T^{(h)} W$ where $T^{(h)}$ is
diagonal with $T^{(h)}_{ss} = e^{i \pi \tau(h,|s|)}$, $|s|$ denotes
the number of 1-bits in $s$ and $\tau$ is another
computationally-efficient real-valued function. With these
definitions, $U^{(h)}_{rs}$ depends only on the distance
$d(r,s)$~\cite{hogg00}, i.e., has the form $U^{(h)}_{rs} =
u^{(h)}_{d(r,s)}$. Quantum computers evaluate this matrix operation
efficiently~\cite{grover96,boyer96,hogg98b}.

Observing the final superposition gives an assignment having $c$
conflicts with probability 
$$ 
p^{(j)}(c)=\sum_{s|c(s)=c} |\psi^{(j)}_s|^2 
$$ 
with the sum over all assignments with $c$
conflicts. In particular, $\Psoln(j) = p^{(j)}(0)$ is the probability
to find a solution in a single trial.

Completing the algorithm requires specifying functional forms for the
phase adjustment functions $\rho$ and $\tau$. Since $c(s)$ and $|s|$
are integers, the matrix elements are unchanged by adding any multiple
of 2 to either $\rho$ or $\tau$. Moreover, changing the sign of both
values simply conjugates the matrix elements. Thus, without loss of
generality, we can restrict the $\rho$ values to be in the range
$(-1,1]$ and $\tau$ in $[0,1]$. In the remainder of this section we
describe the special case equivalent to amplitude amplification and
then discuss one way to include problem structure.

\subsection{Amplitude Amplification}\sectlabel{amplify}

In the notation introduced above, amplitude amplification consists of
the choices
\begin{eqnarray*}
\rho(h,c) &=& \cases{1 & if $c=0$ \cr
                                  0 & otherwise} \\
	& & \\
\tau(h,b) &=& \cases{1 & if $b=0$ \cr
                                  0 & otherwise} \\
\end{eqnarray*}
These choices, which are the same for all steps (i.e., independent of
$h$), cause $P$ to invert the amplitude of solutions and make $U$ a
diffusion matrix with $u_d=-\delta_{d0}+2^{1-n}$ where $\delta_{a b}$
is one if $a=b$ and zero otherwise. Note the off-diagonal elements of
$U$ are exponentially small.

By treating all nonsolution states equally, this algorithm has the
major advantage of a simple expression for the probability to find a
solution after $j$ steps, namely~\cite{boyer96}
\begin{equation}\eqlabel{Psoln amplify}
\Psoln(j) = \sin((2j+1) \theta)^2
\end{equation}
where $\theta = \sin^{-1} \sqrt{S/2^n}$ and $S$ is the number of
solutions. For hard, soluble random $k$-SAT, $2^n \gg S \gg 1$ and $\theta \sim \sqrt{S/2^n}$ is exponentially small. Thus the algorithm can give $\Psoln = \Same{1}$ when $j =
\AtLeast{1/\theta}$, i.e., after an exponentially large number of
steps for hard problems.

In practice, $S$ is not known a priori, so the best choice for the
number of steps $j$ cannot be determined. A useful alternative selects
$j$ differently for each trial as follows~\cite{boyer96}: Starting
with $M=1$,
\begin{itemize}
\item perform a single amplitude amplification trial with the number of steps $j$ selected randomly between 0 and $M-1$
\item if a solution is found, stop. Otherwise, set $M = \min(2^{n/2}, 6M/5)$ and repeat.
\end{itemize}
This procedure increases the expected cost, compared to having prior
knowledge of $S$, by at most a factor of 4~\cite{boyer96}. For the
sake of definite comparison with other choices for $\rho$ and $\tau$,
we describe how to evaluate the expected number of steps to find a
solution.

{
\newcommand{\pRandom}{ {p_{\rm random}} }
\newcommand{\jSoln}{ j_{\rm soln} }
\newcommand{\jNoSoln}{ j_{\rm no\;soln} }
\newcommand{\cost}{ {\rm cost} }

In light of \eq{Psoln amplify}, selecting the number of steps $j$
uniformly at random between 0 and $M-1$, gives the probability to
obtain a solution~\cite{boyer96}
\begin{equation}
\pRandom(M) = \frac{1}{M} \sum_{j=0}^{M-1} \Psoln(j) = \frac{1}{2} - \frac{\sin(4 M \theta)}{4 M \sin(2 \theta)}
\end{equation}
which approaches 1/2 as $M$ increases.

The trial with a given $M$ takes $(M-1)/2$ steps, on average. With
probability $1-\pRandom(M)$ the trial is not successful. Thus the
expected cost for all trials starting with $M$ is
\begin{equation}\eqlabel{amplify cost}
\cost(M) = \frac{M-1}{2} +(1-\pRandom(M)) \; \cost(\min(2^{n/2}, 6M/5))
\end{equation}
When $M \ge 2^{n/2}$, further iterations have $M=2^{n/2}$ so \eq{amplify cost} gives
$$
\cost(2^{n/2}) = \frac{2^{n/2}-1}{2\; \pRandom(2^{n/2})}
$$
For hard, soluble random $k$-SAT, $\pRandom(2^{n/2}) \sim 1/2$ so $\cost(2^{n/2}) \sim 2^{n/2}$.
This condition and \eq{amplify cost} allow computing the expected cost
of the entire loop, i.e., $\cost(1)$, recursively.
%Since $M$ grows exponentially with the number of trials performed, the recursive evaluation of $\cost(1)$ from \eq{amplify cost} requires only $\AtMost{n}$ evaluations.

} % end of group for commands specific to amplitude amplification study

\subsection{Using Problem Structure}

With the algorithm described here, using problem structure is conceptually
straightforward: for a class of problems, such as random $k$-SAT with
given $n$ and $m$, select values for the phase functions $\rho(h,c)$ and
$\tau(h,b)$ to minimize the search cost for typical instances of the class.
We take the number of steps in each trial, $j$, to grow only polynomially with
$n$. Furthermore, $m=\Same{n}$ for hard random $k$-SAT. Thus the number of
values to specify $\rho$ and $\tau$ grows polynomially with $n$. In particular,
for $j$ growing linearly with $n$, $\Same{n^2}$ values completely specify these functions.

We thus have a situation commonly found with developing conventional heuristics: a number of algorithm parameters to tune with respect to the class of problems. Generally, the heuristics are too complicated to permit a useful analytical relation between the parameter values and algorithm cost. Instead, one takes a sample of problem instances and solves them with various choices for the parameter values. Numerical optimization techniques can then find parameter values giving good performance for the sample, e.g., minimizing the median search cost for the sample. These values are evaluated by using them to solve another sample drawn from the same problem ensemble. Since the cost of these heuristics grows exponentially for hard problems, this sampling technique is limited to relatively small problems. Nevertheless, efficient implementations often allow investigating SAT problems with hundreds or thousands of variables.

These remarks also apply to heuristics for quantum computers. On a quantum machine, each trial requires only polynomial time. On the other hand, at least for most parameter choices, $\Psoln$ is exponentially small, thus requiring exponentially many trials to estimate $\Psoln$ on the sample's problem instances. Hence a direct attempt to find parameter values minimizing the median search cost would require exponentially many trials. One way to address this difficulty is to identify how good parameter choices scale with $n$ and then perform the parameter value optimization with smaller $n$. An example of such scaling is having the phase parameters scale as $1/j$, as described below. Another approach makes use of the shift in amplitudes toward low-cost states, illustrated in \sect{behavior}. Thus, instead of maximizing $\Psoln$, we could minimize the expected cost of the state produced by a trial, a quantity easily estimated with a modest number of trials.

Currently, however, such quantum machines do not exist. Instead, we must simulate the quantum algorithm on conventional machines, so each trial requires exponential cost and memory. Thus we are limited to investigating much smaller problems, up to 20 variables or so for SAT. In particular, the simulation evaluates properties of all search states and so is considerably more expensive than evaluating conventional heuristics. The latter, while having exponentially growing costs, typically evaluate only a tiny portion of the full search space.

The number of function evaluations for a numerical optimization procedure grows with the number of values to optimize. Thus as a practical matter,
we consider only a restricted set of possible values with a smaller number of independent parameters. In particular, a study of this algorithm with a fixed number of steps~\cite{hogg98e} suggests restricting $\rho$ and $\tau$ to vary linearly with the number of conflicts $c$ and number of 1-bits $b$, respectively, only slightly reduces the performance. We make this restriction in the specific form for the heuristic presented below.

An alternate approach to finding good parameter values, also discussed below, uses an approximate analytical theory of the algorithm performance. The theory allows rapid evaluation of the approximate performance for a given choice of parameter values. We can then apply numerical optimization to find values giving high performance according to this approximation. This approach allows evaluating behavior for much larger problem sizes, but with the caveat of being only an approximation.

\subsection{Parameter Choices to Use Problem Structure}

To exploit problem structure, we introduce two real-valued functions
$R(\lambda)$ and $T(\lambda)$ defined over $0 \le \lambda \le
1$. These functions specify the amplitude adjustments made,
respectively, by the cost evaluation and mixing, as a function of the
number of steps completed. Specifically, $R$ and $T$ define the phase
adjustment functions as
\begin{eqnarray*}
\rho(h,c) &=& \rho_h c\\
\tau(h,b) &=& \tau_h b
\end{eqnarray*}
with
\begin{eqnarray}\eqlabel{phases}
\rho_h &=& \frac{1}{j} R\left( \frac{h-1}{j} \right) \\
\tau_h &=& \frac{1}{j} T\left( \frac{h-1}{j} \right) \nonumber
\end{eqnarray}
for steps $h=1,\ldots,j$. These values decrease as $1/j$ so $P$ and
$U$ are close to identity matrices as $j$ increases. When iterated
over the $j$ steps of a trial, these operations nevertheless
substantially shift amplitudes among the assignments.  The linearity
of $\tau(h,b)$ with respect to $b$ means the elements of the mixing
matrix are~\cite{hogg00}:
\begin{equation}\eqlabel{mixing}
u^{(h)}_d = \left( e^{i \pi \tau_h/2} \cos( \frac{\pi \tau_h}{2}) \right)^n \left( -i \tan( \frac{\pi \tau_h}{2} ) \right)^d
\end{equation}
Thus the elements decrease rapidly with $d$ so the largest mixing is
among states close to each other.  In the case we consider, $j$ grows
as a power of $n$ (allowing individual trials to complete in
polynomial time). This means the off-diagonal terms of $U$
corresponding to $d=\AtMost{1}$ decrease as a power of $n$ rather than
the exponential decrease of the diffusion mixing matrix.

Completing the algorithm requires explicit forms for $R(\lambda)$ and
$T(\lambda)$ and the number of steps $j$. Ideally these quantities
would minimize the expected total number of steps in all trials for
the particular problem instance. For hard problems, such optimal
choices are not known a priori. Thus we focus instead on functional
forms giving good performance on average for random $k$-SAT, i.e.,
depending only on the ensemble parameters $n$, $k$ and $m$. While the
values could vary from one trial to the next, in analogy with the
procedure described above for amplitude amplification when the number
of solutions is not known, for simplicity we use the same values for
each trial. The expected cost to find a solution is then
$j/\Psoln$. While such choices will not be optimal for each instance,
they can nevertheless improve average performance, as shown below.

\section{Algorithm Behavior}\sectlabel{behavior}

This section illustrates the algorithm's behavior for small problems,
and compares it to amplitude amplification. These observations
motivate the approximate analyses of the following section.

\begin{figure}[t]
\begin{center}
\leavevmode
\epsfbox{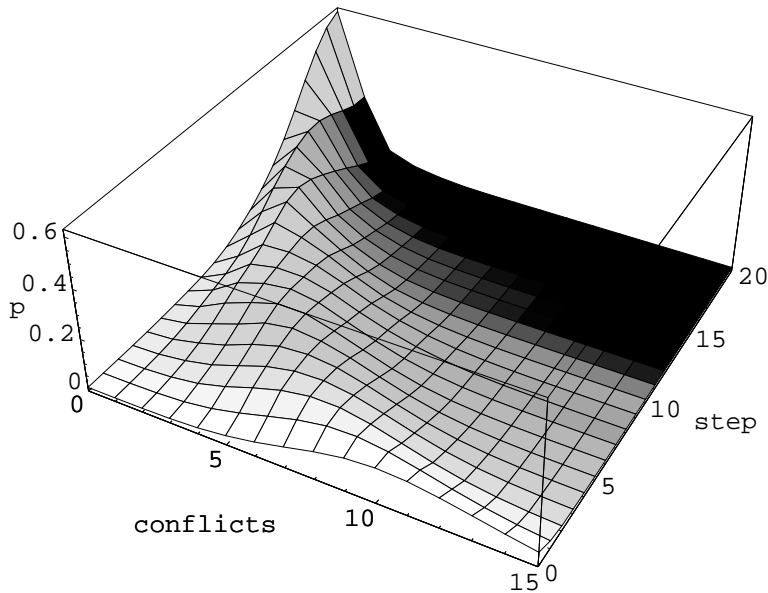}
\end{center}
\caption{\small\label{shift}
Solving a randomly generated 3-SAT problem with $n=20$ and
$\mu=4.25$. For each step $h$, the figure shows the probability
$p^{(h)}(c)$ in assignments with each number of conflicts. Shading is
based on the relative deviations of the amplitudes, described in the
text. The small contributions for assignments with $c>15$ are not
included. This instance has 20 solutions.}
\end{figure}

For $\mu=4.25$, using $j=n$ and linear forms for $R$ and $T$ gives
reasonably good performance. Specifically~\cite{hogg00}, for
\begin{eqnarray}\eqlabel{good parameters}
R(\lambda)	&=&	R_0+R_1(1-\lambda) \\
T(\lambda)	&=&	T_0+T_1(1-\lambda) \nonumber
\end{eqnarray} 
with $R_0=4.86376$, $R_1=-4.18118$, $T_0=1.2$ and $T_1=3.1$,
\fig{shift} shows the behavior for one problem instance. These
numerical values were determined from the approximate analysis, based
on average amplitudes, discussed in \sect{average}.

This figure illustrates several properties of the algorithm.  First,
at each step, probability is concentrated in states with a fairly
small range of costs. Each step shifts the peak in the probability
distribution toward assignments with fewer conflicts, until a large
probability builds up in the solutions. This shift is also seen for
other problem instances (with differing final probabilities) and when
averaged over many samples. The peaks become sharper for larger $n$,
with relative widths decreasing as $\AtMost{1/\sqrt{n}}$. By contrast,
amplitude amplification increases the probability in solutions but all
other amplitudes decrease uniformly.

Second, the variation of amplitudes among states with the same cost is
relatively large only in the last few steps of the algorithm and then
primarily for higher-cost states for which the amplitudes are
small. The shading in \fig{shift} shows this behavior, indicating the
relative deviation of the amplitudes (i.e., ratio of standard
deviation to mean) for states with the each cost, ranging from white
for zero deviation to black for relative deviations greater than 3.

\begin{figure}[t]
\begin{center}
\leavevmode
\epsfig{file=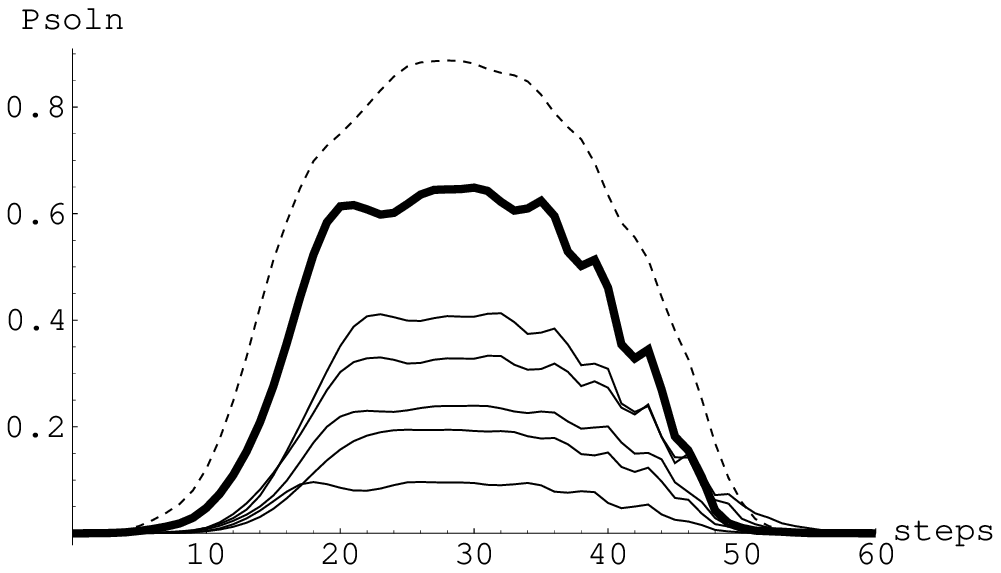,width=2.5in}
\epsfig{file=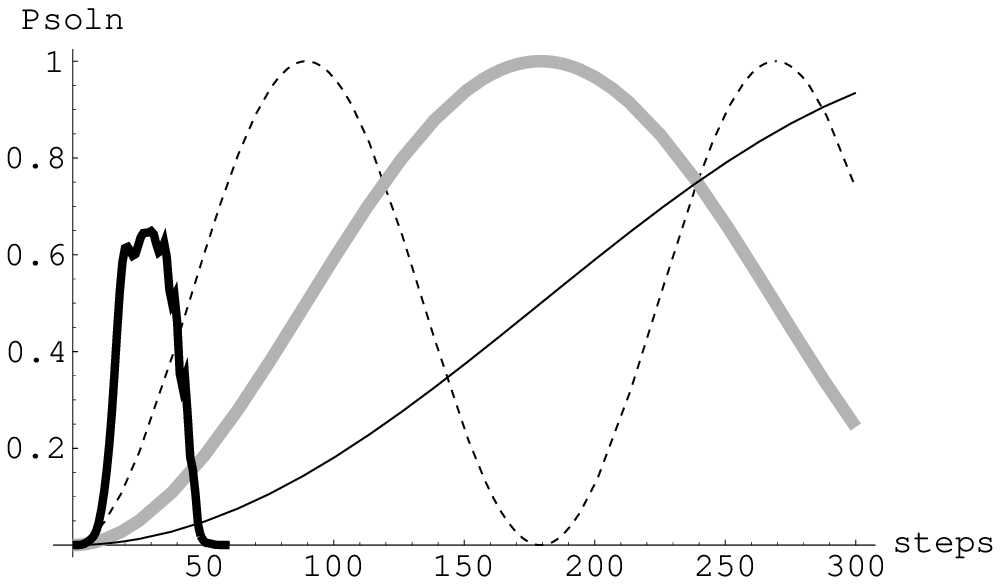,width=2.5in}
\end{center}
\caption[Probability of solution vs.~number of steps.]{\small\label{Psoln vs. steps}
$\Psoln$ as a function of the number of steps for several 3-SAT instances. The first plot shows the behavior of the heuristic algorithm with the same parameters as \fig{shift} and using $j=n$ to define the phase parameters in \eq{phases}. The dashed curve is an instance with 80 solutions, the thick solid curve is the instance with 20 solutions of \fig{shift} and the thin solid curves are different instances with 5 solutions.
The second plot shows the behavior of amplitude amplification with 80, 20 and 5 solutions for the dashed, gray and solid curves, respectively. For comparison, the 20-solution curve from the first plot is also included.}
\end{figure}

\fig{Psoln vs. steps} gives further insight into the algorithm. 
Unlike amplitude amplification, the heuristic reaches its maximum
$\Psoln$ at about the same number of steps for problems with differing
numbers of solutions. Instead, the variation is in the maximum value
of $\Psoln$. Even instances with the same number of solutions behave
differently. Thus for this algorithm, identifying the appropriate
number of steps is not an issue, rather the difficulty is in selecting
appropriate phase parameter adjustments. As problem size increases,
the number of steps required for amplitude amplification increases
exponentially and always gives $\Psoln \approx 1$. By contrast, the
heuristic uses a linearly growing number of steps but $\Psoln$ gets
small.

\begin{figure}[t]
\begin{center}
\leavevmode
\epsfbox{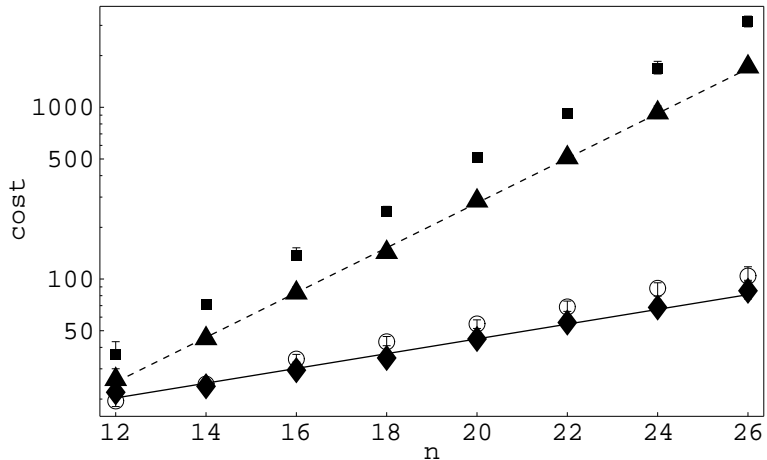}
\end{center}
\caption[Median search costs.]{\small\label{scaling} Log-plot of median search cost
vs.~$n$ for the quantum heuristic (diamond), amplitude amplification
({\em assuming the number of solutions is known} (triangle) or not
(square)), and GSAT~\cite{selman92} with restarts after $2n$ steps
(circle).  For each $n$, the same 1000 soluble random 3-SAT problems
with $\mu=4.25$ were solved with each method (except only 500 and 400
samples for $n=24$ and 26, respectively). For those $n$ not divisible
by 4, half the samples had $m=\lfloor 4.25 n
\rfloor$ and half had $m$ larger by one. Error bars show the 95\%
confidence intervals~\cite[p.~124]{snedecor67}, which in many cases are smaller than the plotted point.  The curves show
exponential fits to the quantum heuristic (solid) and amplitude
amplification (dashed).}
\end{figure}

\begin{figure}[ht]
\begin{center}
\leavevmode
\epsfbox{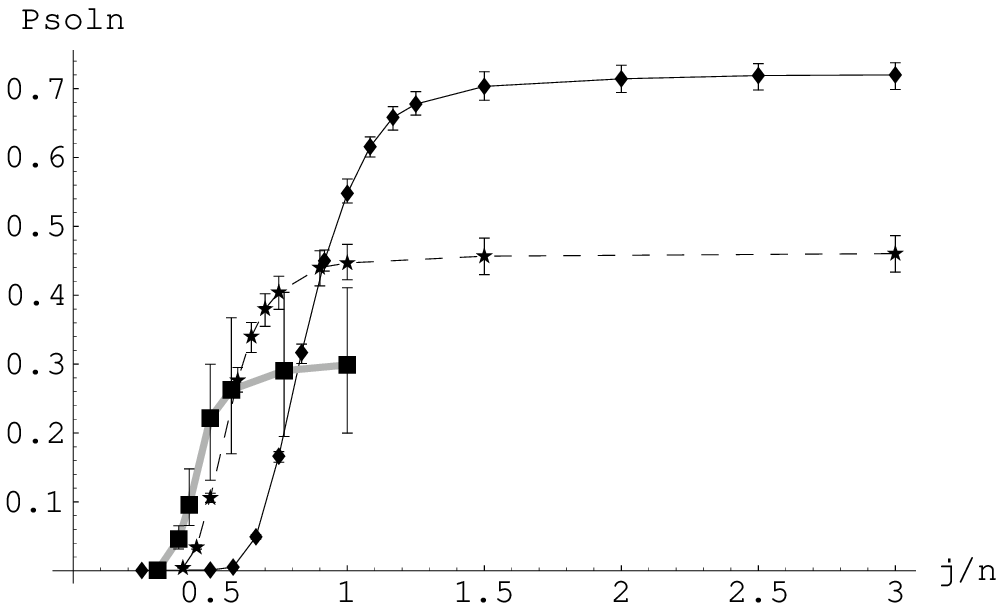}
\end{center}
\caption[Median solution probability vs.~$j$.]{\small\label{steps}Median solution probability
vs.~$j/n$ for the quantum heuristic for 1000 random 3-SAT problems with $n=12$ (solid) and $n=20$ (dashed), and 60 samples with $n=26$ (gray). For these cases $\mu=4.25$.
Error bars show the 95\%
confidence intervals.}
\end{figure}

To compare the net effect of these contrasting behaviors, we examine
the search cost scaling of the two methods.  Using the parameters of
\eq{good parameters}, \fig{scaling} shows the growth of the expected
search cost for randomly generated problems, i.e., the expected number
of steps, $j/\Psoln(j)$. The exponential fit gives the cost growing as
$e^{0.10 n}$. As one caveat, we should note most of the $\Psoln$
values are fairly large for these problem sizes, i.e., $\Psoln \geq
0.3$, thus usually finding a solution after only a few trials. 
Thus much of the variation in costs shown here is due to the
linear growth of the number of steps $j$, and it may require larger
problems to see the cost growth dominated by the behavior of $\Psoln$.

\fig{steps} shows another property of the heuristic: as long as $j/n$ is 
not too small, $\Psoln$ does not change much as $j$ increases. This
can be understood from the scaling of phase parameters of
\eq{phases}. When $j \gg 1$, the algorithm matrices are close to the
identity. In this situation, when $j$ is doubled, the phase
adjustments are halved so two steps change the amplitudes about as
much as the original choice of $j$ did in one step. As a further
observation, the minimum median cost, $j/{\rm median}(\Psoln)$, occurs
at somewhat smaller ratios of $j/n$ as $n$ increases. Exploiting this
decrease gives somewhat lower costs for the quantum heuristic than
those shown in \fig{scaling}, which used $j/n=1$. As a further observation,
the value of $j/n$ giving about half the maximum $\Psoln$ for each $n$
decreases close to linearly on a log-log plot with a slope of about -0.8, indicating the best
scaling performance requires only $j = O(n^{0.2})$. If this behavior
continues to hold for larger $n$, the approximate analysis discussed
below, based on $j \gg \sqrt{n}$, would somewhat overestimate the
minimum possible costs.

Finally we should note the distinction between \fig{Psoln vs. steps}
and \fig{steps}. In the former, the phase parameters of \eq{phases}
are defined using $j=n$ and the behavior of $\Psoln$ is shown for
trials of various numbers of steps using these fixed parameters. In
the latter figure, $j$ varies and gives different phase parameters at
each value of $j/n$, and $\Psoln$ is shown after completing $j$ steps.

\subsection{Comparing with Amplitude Amplification}

Provided the number of solutions $S$ is known, the cost for amplitude
amplification is~\cite{boyer96} $\frac{\pi}{4} \sqrt{2^n/S}$, also
shown in \fig{scaling}. The values grow as $e^{0.30 n}$, i.e., about
three times faster than the quantum heuristic.

In practice, $S$ is not known a priori, requiring the modified
algorithm, described in \sect{amplify}, whose expected cost is less
than four times larger~\cite{boyer96}, so does not affect the
exponential growth rate. However, for the sake of a definite
comparison with the quantum heuristic, which also does not use prior
knowledge of the number of solutions, we compute the actual expected
cost of the modified algorithm using \eq{amplify cost}.  The resulting
values, included in \fig{scaling}, are slightly less than twice as
large as the cost for amplitude amplification when $S$ is known.

\subsection{Comparing with a Conventional Heuristic}

Average costs for even the best known classical heuristics grow
exponentially. For instance, \fig{scaling} shows the search cost for a
good classical heuristic, GSAT~\cite{selman92}, grows slightly faster
than this quantum heuristic.  The GSAT algorithm starts from a random
assignment and, for each step, examines the number of conflicts in the
assignment's neighbors (i.e., assignments obtained by changing the
value for a single variable) and moves to a neighbor with the fewest
conflicts. If a solution isn't found after a prespecified number of
steps, e.g., because the current assignment is a local minimum, the
search is tried again from a new random assignment. The most
significant comparison between GSAT and the quantum heuristic is the
relative growth rates in the search costs, as measured by the number
of steps. The corresponding actual search times will depend on detailed
implementations of the steps. Although the number of elementary
computational steps, involving evaluating the number of conflicts in
an assignment (and, in the case of GSAT, its neighbors) are similar
for both techniques, differences in the extent to which operations can
be optimized away (e.g., as is possible in some cases for NMR-based
quantum implementations~\cite{chuang98}) and the relative clock rates
of classical and quantum machines remain to be seen. 

At any rate, the figure shows that including the number of conflicts
in the phase adjustments reduces the number of steps required for the
quantum algorithm below that required for GSAT, on average. Because
the trials are independent, both the quantum heuristic introduced here
and GSAT can be quadratically improved with amplitude
amplification~\cite{brassard98}, amounting to decreasing the growth
rates shown in the figure by a factor of two.  Such an improvement
requires extending coherence across multiple trials, rather than just
a single one. 

This technique also generalizes to allow the quantum heuristic presented here to work with the results of deterministic classical heuristics with independent trials (e.g., a deterministic version of GSAT in which, say, any ties are broken by selecting the first neighbor with minimum cost in a lexicographic ordering of the states, or the seed used with the random number generator is prespecified for all trials). Specifically,
instead of basing the phase adjustment on the number of conflicts in
a state, we run GSAT starting from that state for a fixed number
of steps. We can then use the number of conflicts of the resulting state
as the basis for the phase adjustment. This thus uses more information
about the heuristic than just combining it with amplitude amplification, which tests whether the heuristic finds a solution~\cite{brassard98}. For the problem sizes discussed here, using this technique gives considerably higher probabilities to find a solution, even using the same phase adjustment parameters as used for the original method involving the number of conflicts in the states. However, the additional steps required to evaluate GSAT within each trial, results in a larger overall cost. Nevertheless, this technique may be useful for larger problem sizes, where the probabilities to find solutions are lower.

From this discussion, the structured quantum algorithm appears to
improve on GSAT for hard SAT problems, but definitive statements
cannot be made based only on such small problems.  Unfortunately,
classical simulations of quantum machines incur an exponential growth
in time and memory, preventing evaluation with larger problems. More
extensive empirical evaluation requires either faster simulation
techniques, perhaps approximate~\cite{cerf97}, or quantum computers.

%%%%%%%%%%%%%%%%%%%%%%%%%%%%%%%%%%%%%%%%%%%%%%%%%%%%%%%%%%%%%%%%
\section{Approximate Analyses of Behavior}

The usual approach to evaluating conventional heuristics, and tuning
any adjustable parameters they may have, is by running them on a
sample of problems. This is necessary because analytical methods are
often unable to account for the complicated dependencies in the search
path explored by the heuristic. As discussed in the previous section,
such simulations are also useful for quantum algorithms, but are
limited to small problems.

As a complementary approach, we consider approximate analytical
techniques. The average properties of random $k$-SAT successfully help
understand and improve search methods, both
classical~\cite{edelkamp98,cheeseman91,hogg95c,gent96} and
quantum~\cite{hogg00}. In particular, the quantum algorithm operates
with the entire search space at each step so its performance depends
on averaged properties of the search states. For simple ensembles,
such as random $k$-SAT, such averages are readily computable and thus
give asymptotic characterizations of the problems for large $n$.  In
addition to estimating algorithm performance, such analyses provide
insight into the qualitative features of the behavior seen
empirically.

For a problem instance $P$, let $\psi^{(h)}_s(P)$ be the amplitude for
state $s$ after completing step $h$ of the algorithm. Initially,
$\psi^{(0)}_s(P)=2^{-n/2}$. A single step of the algorithm, from
\eq{map}, gives
\begin{equation}\eqlabel{step}
\psi^{(h)}_r(P) = \sum_s u^{(h)}_{d(r,s)} e^{i \pi
\rho_h c(s)} \psi^{(h-1)}_s(P)
\end{equation}
The remainder of this section discusses techniques using the
properties of random $k$-SAT to estimate the algorithm's behavior for
large $n$, and suggest suitable choices for the phase functions.

\subsection{Average $\Psoln$}

Ideally, we would like to estimate the typical search cost for
problems in the ensemble. The expected cost for a given problem
instance is $j/\Psoln$. Thus one quantity to examine is the
ensemble-average $\expect{j/\Psoln}$ or, since, in the case considered
here, $j$ is the same for all instances, $j
\expect{1/\Psoln}$. However, this quantity is infinite if even a
single problem instance is insoluble. Even restricting attention to
soluble instances, we find a wide variation in solution costs. Thus
the average is dominated by a small fraction of the instances and does
not indicate typical behavior. A better indication is the median value
of $j/\Psoln$, but is difficult to treat analytically. As an
analytically tractable quantity, we focus instead on
$j/\expect{\Psoln}$.

The random $k$-SAT ensemble includes both soluble and insoluble
instances, so $\expect{\Psoln} \leq \Psoluble$, the fraction of
soluble instances in the ensemble.  Below the phase transition, near
$\mu=4.25$ for random 3-SAT, $\Psoluble \rightarrow 1$.  For larger
$\mu$, $\Psoluble \rightarrow 0$ as $n$ increases, in which case the
performance for {\em soluble} problems is given instead by
$\expect{\Psoln}/\Psoluble$.  Unfortunately, the random $k$-SAT
ensemble does not have a simple expression for $\Psoluble$, or even
just its leading exponential scaling rate, precluding an exact
evaluation for overconstrained soluble problems.  One approach to
estimate this behavior uses empirical classical search to evaluate
$\Psoluble$ for a range of problem sizes for a given value of
$\mu$. The behavior of these values as a function of $n$ then
estimates the scaling of $\Psoluble$. For instance, samples of $10^4$
problems for $n$ from 50 to 250 show~\cite{selman95} close to
exponential decrease of $\Psoluble$ for $\mu$ values somewhat above
the transition. The resulting estimates of the actual decay rates for
$\Psoluble$ are $0.011$, $0.025$ and $0.045$ for $\mu$ equal to $4.5$,
$4.7$ and $4.9$, respectively. Analytic bounds on the behavior of
$\Psoluble$ for $\mu$ between 4.2 and 5.2 are difficult to obtain. One
such result is $\Psoluble \leq \exp(-5.9 \times 10^{-5} n)$ at
$\mu=4.762$~\cite{kamath94}, which is a considerably smaller decay
rate than suggested by empirical evaluation. Above $\mu=5.2$, the
expected number of solutions, $\expect{S}$, goes to zero, so the
Markov bound $\Psoluble \leq \expect{S}$ provides another constraint.

For random $k$-SAT $\expect{\Psoln}$ is
\begin{equation}
\frac{1}{\Nclauses^m} \sum_P \sum_{s | c(s)=0} | \psi^{(j)}_s(P) |^2
\end{equation}
where the outer sum is over the $\Nclauses^m$ possible problem
instances and the inner sum is over those assignments $s$ that are
solutions for $P$. Since the clauses are selected independently, the
sum over problems is equivalent to $m$ sums, each of which ranges over
$\Nclauses$ possible clauses.  Interchanging the order of summation
gives an outer sum over all assignments $s$ and an inner sum over
those problem instances for which $s$ is a solution, i.e., instances
containing no clause conflicting with $s$. Since the random $k$-SAT
ensemble treats all assignments equally, this sum over problems is the
same for all choices of $s$. Thus we can focus on a single assignment,
say $s=0 \ldots 0$. For an assignment $r$ and clause $\sigma$, let
$\alpha(r,\sigma)=1$ if $\sigma$ conflicts with $r$ and otherwise is
zero. Then $c(r)$ for problem instance $P$ is the sum of
$\alpha(r,\sigma)$ over the clauses in $P$.  With this notation,
\eq{map} gives $\expect{\Psoln}$ equal to
\begin{equation}
%\sum_{s_0,\ldots,s_{j-1}} \sum_{s'_0,\ldots,s'_{j-1}}
\sum_{	\begin{array}{c}
		\scriptstyle s_0,\ldots,s_{j-1}	\\
		\scriptstyle s'_0,\ldots,s'_{j-1}
	\end{array}} 
	\left( \prod_{h=1}^j \upsilon_h {\upsilon'_h}^{*} \right)
	\left(
		\frac{1}{\Nclauses} \sum_{\sigma} 
			\exp \left( 
			i \pi \sum_h \rho_h (\alpha_h - \alpha'_h )
			\right)
	\right)^m
\end{equation}
where $\upsilon_h \equiv u^{(h)}_{d(s_h,s_{h-1})}$, $\upsilon'_h \equiv u^{(h)}_{d(s'_h,s'_{h-1})}$, $\alpha_h \equiv \alpha(s_{h-1},\sigma)$, $\alpha'_h \equiv \alpha(s'_{h-1},\sigma)$ and we define $s_j \equiv s'_j \equiv s$.
The $\sigma$ sum is over all clauses not conflicting with $s=0 \ldots 0$, i.e., those $\sigma$ with $\alpha(s,\sigma)=0$.
% $d_h \equiv d(s_h,s_{h-1})$, $d'_h \equiv d(s'_h,s'_{h-1})$

\begin{figure}[t]
\begin{center}
\leavevmode
\epsfbox{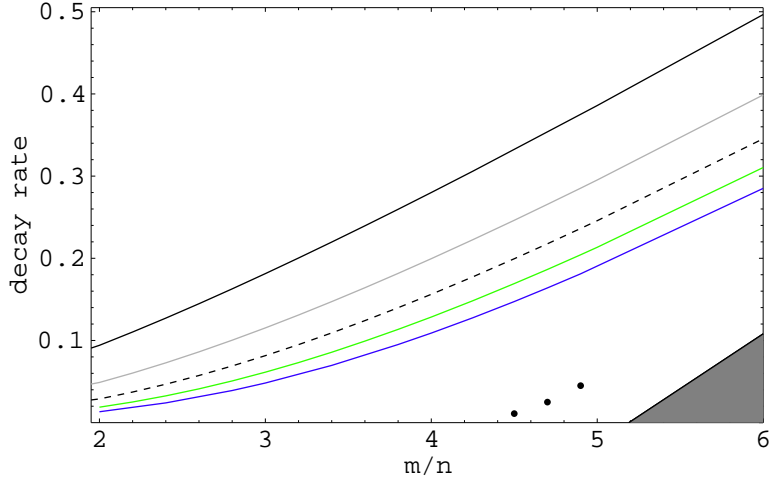}
\end{center}
\caption[Constant number of steps.]{\small\label{multi}Minimum decay rates for  $\expect{\Psoln}$ as a function of $\mu=m/n$ for (from top to bottom) $j=1$ through 5 steps using linear variation in phase parameters with step number (i.e., of the form given in \eq{good parameters}), but with different numerical values for each $j$. The points indicate empirical estimates of the decay rate for $\Psoluble$, a lower bound on the decay rate for $\Psoln$. The upper edge of the filled region is, in turn, a lower limit on $\Psoluble$ given by the Markov bound using the expected number of solutions.}
\end{figure}

For constant $j$, an exact asymptotic analysis~\cite{hogg98e} shows
$\expect{\Psoln}$ decays exponentially but at smaller rates than the
exponential growth in number of steps required by amplitude
amplification. \fig{multi} shows examples for $j$ up to 5:
specifically the decay rate $A$ defined by $\expect{\Psoln} \sim
\exp(-A n)$.
For example, with $\mu=4.25$, for $j=5$ the decay rate is $A=0.13$,
only slightly larger than the empirical growth rate of the median cost
shown in \fig{scaling} when $j=n$. This may indicate the problem sizes
feasible to simulate are not large enough to show the full benefit of
allowing $j$ to grow with $n$.

This analysis is useful in suggesting the scaling behavior of
\eq{phases} for the algorithm's parameters and shows $\expect{\Psoln}$
decreases less rapidly as $j$ increases. Unfortunately, this analysis
is not applicable to the more interesting situation where the number
of steps $j$ grows with $n$. While it may be possible to develop
approximations for $j \gg 1$, a simpler approach uses the observed
properties of the amplitudes seen in \sect{behavior}. This approach is
described in the remainder of this section.

\subsection{Average Amplitudes\sectlabel{average}}

\begin{figure}
\begin{center}
\begin{picture}(100,150)
	\put(0,8){\line(1,0){100}}
	\put(50,0){\makebox(0,0)[t]{$c$}}
	\put(0,0){\qbezier(20,10)(50,200)(80,10)}
	\put(80,40){\makebox(0,0)[l]{$v(c)$}}
	\put(0,0){\qbezier(0,100)(30,30)(60,10)}
	\put(0,113){\makebox(0,0)[t]{$|A_c|^2$}}
	\thicklines
	\put(0,0){\qbezier(10,15)(30,250)(50,15)}
	\put(40,122){\makebox(0,0)[l]{$v(c) |A_c|^2$}}
\end{picture}
\end{center}
\caption{\small\label{expansion}Schematic behavior of average amplitudes, on a logarithmic scale, as a function of number of conflicts $c$. The average number of states with $c$ conflicts, $v(c)$, is sharply peaked around the average number of conflicts $m/2^k$. When the magnitude of the amplitudes decreases rapidly with $c$, as shown here, the probability in states with $c$ conflicts is also sharply peaked, but at a somewhat lower value, corresponding to the shift seen in \fig{shift}. Quantitatively, the values decrease exponentially with $n$, so the logarithms, shown here, are proportional to $n$ and the relative width of each peak is $\AtMost{1/\sqrt{n}}$.}
\end{figure}
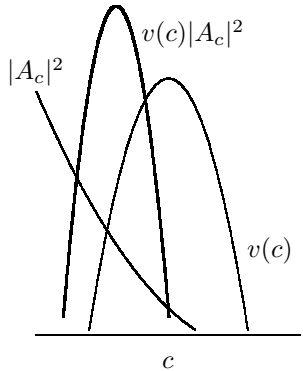

Empirical evaluations show that amplitudes for states with the same
number of conflicts are generally quite similar.  This observation
motivates an analysis based on the behavior of the average amplitude
for states with each cost~\cite{hogg00}. Consider the quantity
$A^{(h)}_C$ defined as $\expect{\psi^{(h)}_s(P)}$ with the average
first over all states $s$ with $c(s)=C$ in the problem instance $P$,
and then over all problems in the random $k$-SAT ensemble with given
$n$ and $m$. Assuming amplitudes for states with the same cost are the
same, at least for the dominant cost states at each step, \eq{step}
becomes
\begin{equation}\eqlabel{avg}
A^{(h)}_C \approx \sum_{d,c} u^{(h)}_d e^{i \pi \rho_h c} \states{d}{C,c} A^{(h-1)}_c
\end{equation}
where $\states{d}{C,c}$ is the expected number of states with $c$
conflicts at distance $d$ from a state with $C$
conflicts. Significantly, $\states{d}{C,c}$ is a property of the
problem ensemble, independent of the algorithm details. For random
$k$-SAT, $\states{d}{C,c}$ is a multinomial sum described briefly in
the appendix.

Simulations show probability concentrates in a small range of cost
values, as illustrated schematically in \fig{expansion}. These
dominant costs are, in turn, close to the average cost $\sum_C C v(C)
|A_C|^2$ where $v(C)$ is the expected value, for random $k$-SAT, of
the number of states with $C$ conflicts. We can thus expand $A_c
\approx A_C Z^{c-C}$, around the average cost, with $Z$ a complex
number depending on the step.
%This expansion then allows direct evaluation of the sum over $c$ in \eq{avg}.

In one step, \eq{avg} implies $Z$ changes by $\AtMost{1/j}$. So for
$j \gg 1$, $Z$ becomes a smooth function of $\lambda=h/j$ satisfying the differential equation~\cite{hogg00}
\begin{equation}\eqlabel{avg deqn}
Z_{\lambda} = i \pi Z \left( R - \frac{T}{2} k f \frac{(1-p(1-Z))(1-Z)}{(1-p)Z} \right) 
\end{equation}
where $\chi = \frac{|Z|^2 p}{1-p(1-|Z|^2)}$ and
$$
f = \exp \left(-k \mu  (1-Z) \left( \frac{p(1-\chi)}{1-p}-\frac{\chi}{Z}  \right) \right)
$$
The initial condition, corresponding to all amplitudes equal, is $Z(0)=1$.
With this approximation, the dominant cost value is $\chi m$.

With suitable choices for $R$ and $T$, such as those in \eq{good
parameters} for $k=3, \mu=4.25$, \eq{avg deqn} gives $Z(1)=0$ thereby
predicting most of the amplitude concentrates in states with the
fewest conflicts, i.e., solutions if the problem instance is soluble.

\subsection{Including Variation Among Amplitudes}

The approximation based on average amplitudes shows good
correspondence with empirical evaluation for most of the steps of the
algorithm. However the variation among amplitudes with the same costs
increases for the last few steps of the algorithm, as illustrated in
\fig{shift}. If this variation remains significant as problem sizes
increase, especially among states with fairly low costs, it remains
possible that the small averages predicted for nonsolution states when
$Z(1)=0$ are due to large variation in the phases of the amplitudes
rather than small magnitudes, leading to somewhat less concentration
in solution states than predicted.

It is thus useful to estimate the contribution from this variation to
the behavior of the algorithm. A direct approach would consider the
ensemble-average of the variance in amplitudes among states with each
cost. Such an analysis gives a similar prediction, namely appropriate
phase functions can concentrate amplitude sufficiently into low-cost
states to give high average performance. To avoid introducing
significant variation in amplitudes for states with the same cost, the
resulting phase adjustments are smaller than those based on the
behavior of the average amplitudes alone. Using such parameters for
small problems gives significantly lower $\Psoln$ values, and hence
higher costs, than shown in \fig{scaling}. Since the analysis assumes
$\sqrt{n} \gg 1$, this poor performance for small $n$ could be due to
the small problem sizes.

Another possibility is the analysis based on the variance of
amplitudes overestimates the effect of amplitude
variation. Specifically, in the case treated here, where the number of
steps grows with $n$, most contribution to the amplitude of a given
state is from other states relatively near to it. This is due to the
decreasing values of $\tau_h$ in \eq{phases} causing the mixing matrix
elements $u_d$ in \eq{mixing} to decrease rapidly with distance. Thus
large variations among amplitudes for states that are far apart do not
much affect the result of a single step. Conversely, nearby states
share many of the same neighbors so their amplitudes are likely to be
more correlated than those of distant states. This means a useful
characterization of the amplitude variations should also account for
the distance between the states. Thus we consider an approximation
based on the assumption that {\em nearby} states with the same costs
have approximately the same amplitudes.

\begin{figure}[t]
\begin{center}
\begin{picture}(100,120)
% show relations among states r,r',s,s'
\put(0,0){\framebox(20,20){$r$}}
\put(80,0){\framebox(20,20){$r'$}}
\put(0,80){\framebox(20,20){$s$}}
\put(80,80){\framebox(20,20){$s'$}}
\put(10,20){\line(0,1){60}} \put(0,50){\makebox(0,0){$d$}}
\put(90,20){\line(0,1){60}} \put(100,50){\makebox(0,0){$d'$}}
\put(20,10){\line(1,0){60}} \put(50,0){\makebox(0,0){$D$}}
\put(20,90){\line(1,0){60}} \put(50,100){\makebox(0,0){$\delta$}}
\put(-7,-7){\makebox(0,0){$C$}}
\put(107,-7){\makebox(0,0){$C'$}}
\put(-7,107){\makebox(0,0){$c$}}
\put(107,107){\makebox(0,0){$c'$}}
\end{picture}
\end{center}
\caption{\small\label{states}Distance relations and costs for the four assignments $r$, $r'$, $s$ and $s'$.}
\end{figure}
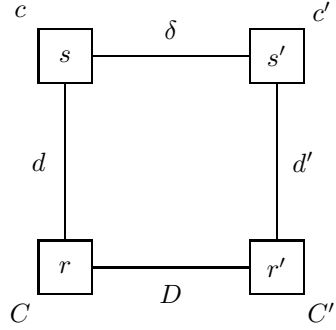

Consider the quantity $S^{(h)}_{D,C,C'}$ defined as
$\expect{\psi^{(h)}_s(P) \psi^{(h)}_{s'}(P)}$ with the average first
over all pairs of states $s,s'$ such that $d(s,s')=D$ and $c(s)=C$,
$c(s')=C'$ in the problem instance $P$, and then over all problems in
the random $k$-SAT ensemble with given $n$ and $m$.  A mean-field
approximation with \eq{step} gives
\begin{equation}\eqlabel{S mf}
S^{(h)}_{D,C,C'} \approx  \sum_{d,d',\delta,c,c'} u^{(h)}_d u^{(h)*}_{d'} e^{i \pi \rho_h (c-c')} \states{D,d,d',\delta}{C,C',c,c'} S^{(h-1)}_{\delta,c,c'}
\end{equation}
where $\states{D,d,d',\delta}{C,C',c,c'}$ is the ensemble average of
the number of assignment pairs $s,s'$ with costs $c,c'$, respectively, with
distance relations $d(s,s')=\delta$, $d(r,s)=d$, $d(r',s')=d$,
averaged over all assignment pairs $r,r'$ with $d(r,r')=D$ and costs $C,C'$,
respectively, as illustrated in \fig{states}.

\eq{S mf} uses $\states{D,d,d',\delta}{C,C',c,c'}$, which characterizes the 
relevant structure of the problems and is independent of the search
algorithm choices for $R(\lambda)$ and $T(\lambda)$. As with
$\states{d}{C,c}$, this 4-state structure quantity is a sum of
products of multinomials. As described in the
appendix, an expansion similar to that described above for the average
amplitude $A_C$ gives an asymptotic expansion of \eq{S mf} for large
problem sizes. Specifically, we express the behavior for $S$ with the
expansion near $D=0$, $c \approx C$, $d \approx D$ of $S \propto Y^d
X^c$, with $X \equiv r e^{i \theta}$ and $Y$ depending on the
step. For $j \gg 1$, these values change slowly from one step to the
next giving differential equations:
\begin{eqnarray}\eqlabel{deqn}
Y_{\lambda} 	&=& \pi T \left(
	Y^2 k \mu F \frac{\nu}{1-p} {\left( 1 - r \right) \,
  \left( 1 + p\,\left( -1 + k\,r \right)  \right) \,
  \sin (\theta )} + G \sin(B)
	\right) \\ \nonumber
r_{\lambda} 	&=& -\frac{\pi T}{2} k Y F \frac{\nu}{1-p} \sin(\theta) \\
\theta_{\lambda}&=& \pi R - \frac{\pi T}{2 r} \left(
		G \frac{1}{Y} ( \cos(B-\theta) - r^2 \cos(B+\theta) )
		- F Y (k-1) (r-\cos(\theta))
	\right)	\nonumber
\end{eqnarray}
with
\begin{eqnarray*}
F	&=&	\exp \left( 
	-\nu k \mu {\left( 1 + {r^2} - 2\,r\,\cos (\theta ) \right)} 
	\right) \\
G	&=&	\exp \left( 
	\nu { k\,\mu \,\left( \left( 1 + {r^2} \right) \,\cos (\theta ) -2\,r \right)   } 
	\right) \\
B	&=&	\nu {k\,\mu \left( -1 + {r^2} \right)  \sin(\theta )}
\end{eqnarray*}
and $\nu = \frac{p}{1-p(1-r^2)}$.  The initial conditions are
$r(0)=1$, $\theta(0)=0$ and $Y(0)=1$, corresponding to all amplitudes
equal.  The equations for $Y$ and $r$ are unchanged by adding any
multiple of $2 \pi$ to $\theta$.

\subsection{Predicted Behavior Using Amplitude Variation}

\begin{figure}[t]
\begin{center}
\leavevmode
\epsfbox{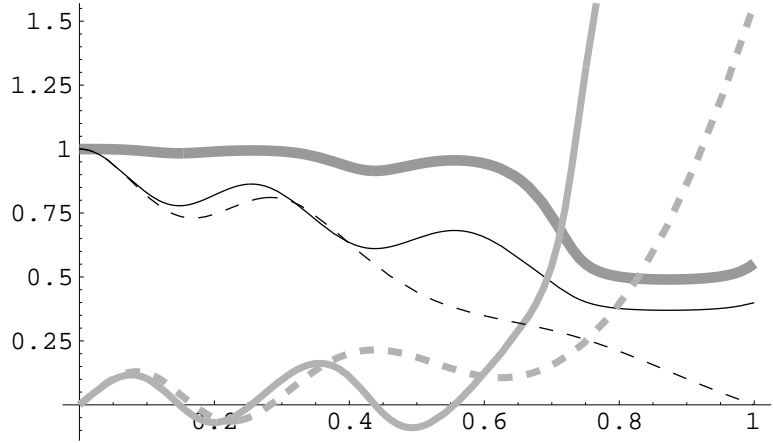}
\end{center}
\caption[Approximation behaviors.]{\small\label{deqn}Behavior of solutions to approximations as a function of $\lambda=h/j$, with the same phase parameters used in \fig{scaling}. Solid curves are from \eq{deqn} and dashed from \eq{avg deqn}. The black curves are $r=|X|$ (solid) and $|Z|$ (dashed) while the light gray curves are the arguments of $X$  (i.e., $\theta$) and $Z$. The thick dark gray curve is $Y$.}
\end{figure}

\fig{deqn} shows the solution of \eq{avg deqn} and \eq{deqn} for the choice 
of $R$ and $T$ of \fig{shift}. For these parameters \eq{avg deqn}
gives $|Z| \rightarrow 0$ predicting good performance. But \eq{deqn}
has $|X| \equiv r$ remaining positive. For small $\lambda$, the variation among
amplitudes for states with the same cost is small, so
$S^{(h)}_{D,C,C'} \approx A^{(h)}_C A^{(h)*}_{C'}$, corresponding to $X \approx
Z$. As $\lambda$ increases, these quantities differ significantly so
the two approximations make quite different predictions for the form
of the amplitudes at the end of the trial, i.e., at $\lambda=1$.

As one quantitative evaluation of the approximation including
amplitude variation, we can compare its prediction of the scaling of
$\Psoln$ with that seen in \fig{scaling}. This approximation has
$\expect{|\psi_s|^2} \propto |X|^{2c}$ for states with $c$
conflicts. Thus, assuming this expansion holds not only for dominant
$c$ but also extends to $c=0$, i.e., solutions, the probability for a
solution, as described in \eq{prob soln} of the appendix, is 
$$
\left( \frac{1-p}{1-p(1-|X|^2)} \right)^m
$$
with $p=2^{-k}$. The solution to \eq{deqn} shown in \fig{deqn} has
$|X|\equiv r =0.399$ at $\lambda=1$, giving $\expect{\Psoln} \sim
\exp(-0.0957 n)$ since $m = 4.25 n$, and thus estimates the cost
$j/\expect{\Psoln}$ growing as $\exp(0.0957 n)$, very close to the
observed growth of $e^{0.10 n}$ in \fig{scaling}. Note the latter
quantity is based on median costs of soluble problems while the theory
is an estimate of $j/\expect{\Psoln}$ for all problems.

\fig{deqn} shows the contribution from the spread remains small, i.e., $Y$ is near 1,
for most of the steps, and then decreases. This corresponds to
empirical observations of the behavior of the algorithm where
amplitudes for states near the dominant cost have relatively little
variation until the last few steps~\cite{hogg00}, as illustrated in
\fig{shift}. Thus \eq{deqn} provides an account of this behavior.

These observations show the relations among groups of four assignments
in random SAT problems, used to derive \eq{deqn}, give a fuller
account of the algorithm behavior than the simpler theory of the
average amplitudes. It does not, however, account for all
behaviors. For example, as shown in \fig{Psoln vs. steps}, continuing
the trial beyond step $j$, i.e., for $\lambda>1$, gives small
oscillations in $\Psoln$ up to a bit below $\lambda=2$ followed by a
drop to $\Psoln \approx 0$. On the other hand, continuing the solution
of \eq{deqn} beyond $\lambda=1$ gives small oscillations in $r$ even
beyond $\lambda=3$. That is, \eq{deqn} fails to account for the drop
in $\Psoln$ beyond $\lambda \approx 2$ for these parameters. Examining
the amplitudes shows they develop multiple peaks so there is no longer
a single small range of dominant costs as assumed in deriving
\eq{deqn} from \eq{S mf}. Nevertheless, \eq{deqn} appears reasonable
for describing the behavior over the range of most value for finding
solutions, i.e., the range over which the bulk of the amplitude
concentrates in low-cost states.

Of particular interest is whether this approximation can also suggest
improvements to the algorithm, i.e., the choices of the functions
$R(\lambda)$ and $T(\lambda)$.  As illustrated in \fig{deqn}, the
solutions to \eq{deqn} take on finite nonzero values after the final
step, at $\lambda=1$, for most choices of the phase
functions. However, there are special cases in which
$r(1)=0$.

To see what this requires, note that $r_{\lambda}$ in \eq{deqn} is
proportional to $Y$. Thus for $r$ to decrease to zero, it is important
to prevent $Y$ from also becoming small too rapidly. Examining the
right-hand sides of \eq{deqn} shows $Y$ decreases much more rapidly
than $r$, when $r$ is small, unless $\theta$ is near $\pi$. Thus one
way to have $r(1)=0$ is for $\theta \rightarrow \pi$ while $Y$ remains
bounded above zero. In this case, the $1/r$ term contributing to
$\theta_\lambda$ in \eq{deqn} becomes large. Thus if $\theta$ is to
approach $\pi$ smoothly, the phase adjustment $R(\lambda)$ must also
be large near $\lambda=1$. In particular, the linear form of
\eq{good parameters} near $\lambda=1$ is not be sufficient to allow
$r \rightarrow 0$.

Solutions of \eq{deqn} with $r(1)=0$ will not, in general, also
satisfy the initial conditions $r(0)=1$, $\theta(0)=0$ and
$Y(0)=1$. Nevertheless, appropriate choices of the phase parameter
functions, $R(\lambda)$ and $T(\lambda)$, satisfy both sets of conditions. These choices can be found numerically using
parameterized forms for these functions and adjusting the parameters
to match the required conditions.  In these cases $r(\lambda) =
\Same{1-\lambda}$ near $\lambda=1$. So for finite $n$, when
$\lambda=1-\AtMost{1/j}$, i.e., the last few steps, we have
$r=\AtMost{1/j}$.  Thus this approximate analysis indicates $j/\expect{\Psoln}$ is polynomial in $n$, hence predicting high
performance is possible when the number of steps is much greater than
$\sqrt{n}$.

The choices for the phase parameters are not unique. The additional
flexibility may be useful to minimize the variation in performance
among different problem instances in the ensemble. More significantly,
the need for large phase adjustments for the last few steps to have $r
\rightarrow 0$ suggests the asymptotic character of the algorithm may
change in the last steps of a trial. In particular, the large
adjustments may mean the differential equations of \eq{deqn} are no
longer good approximations for the discrete map \eq{S mf}, thereby requiring a more detailed asymptotic analysis for the behavior in the last few steps.
In particular, this observation highlights two distinct approximations:
first replacing \eq{step} by \eq{S mf} and then estimating its asymptotic
behavior by a system of differential equations in \eq{deqn}. The
latter approximation depends on relatively small changes from one
step to the next~\cite{wormald95} which no longer holds when using
large phase parameters.

\begin{figure}[t]
\begin{center}
\leavevmode
\epsfbox{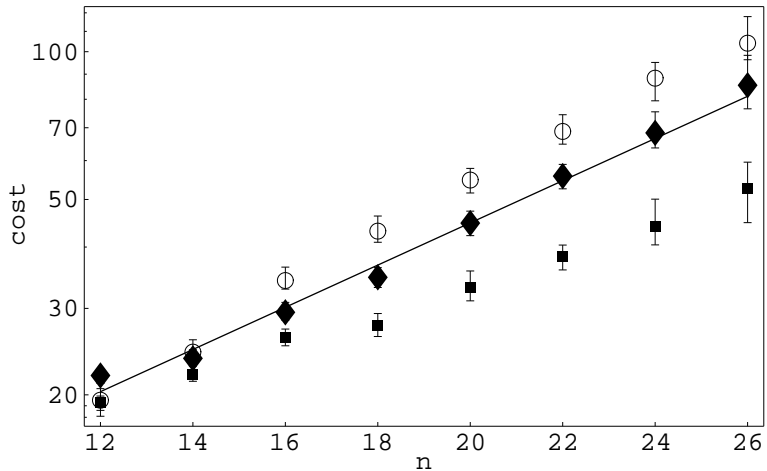}
\end{center}
\caption[Cost scaling with adjusted parameters.]{\small\label{last}
Search cost scaling with the number of steps and phase parameters for the last two steps adjusted to improve performance (square). For comparison, the plot includes the original heuristic (diamond) and GSAT (circle) values from \fig{scaling}. For each $n$, all search methods use the same problem instances.}
\end{figure}

Empirically, using the large $R$ values near $\lambda = 1$ required by this analysis gives small values for $\Psoln$ for the small problems feasible to simulate when $j \approx n$. However, when the number of steps $j$ is taken quite large, e.g., $j \approx 100$ for $n=14$, large $R$ values for the last few steps do give large $\Psoln$, but then the cost $j/\Psoln$ is large due to the large number of steps. Thus a good test of this theory's predictions is beyond the range of feasible simulation. 

Nevertheless, the analysis indicates a change in behavior for the last few steps so we consider separate choices for $\rho_h$ and $\tau_h$ for the last few steps. That is, we numerically optimize the values separately for the last few steps while continuing to use the linear variation of \eq{phases} for the remaining steps. For example, \fig{last} shows the performance using separate values for the last two steps and reducing $j/n$ to minimize median cost as suggested by \fig{steps}. The figure shows improved performance with these adjustments, i.e., using a slower growth of $j$, e.g., $j = \Same{n^{0.2}}$, and different phase adjustments for the last two steps. An exponential fit to the new values gives the median cost growing as $e^{0.08 n}$, a somewhat smaller growth rate than the original heuristic.

Because the numerical optimization requires solving a sample of problems multiple times, finding good values for $\rho$ and $\tau$ is only possible with small samples, e.g., 50 instances with $n=16$, 10 with $n=22$ and 1 with $n=24$. For $n=26$, parameter optimization is not feasible so the figure shows the behavior using the parameters found for $n=24$. Thus the resulting optimal phase values for these small samples are not likely to be the best possible for the ensemble as a whole. Hence the reduced median costs shown in \fig{last} are upper bounds on the possible performance of the heuristic for these problem sizes. A more comprehensive evaluation would optimize phase parameters separately for every step and use larger training samples. Such a procedure is only feasible for even smaller problems than shown in the figure. Nevertheless, it appears likely that linear variation in the phase parameters is quite good for all but the last few steps.

\section{Discussion}

As we have seen, search state properties are readily incorporated in
quantum search algorithms through amplitude phase
adjustments. Properly selected, such adjustments achieve lower overall
cost than unstructured search, and require less coherence time for the
quantum operations. On the other hand, the additional complexity of
such algorithms precludes a simple analytic expression of their
average cost and hence makes it difficult to identify those phase
choices giving the minimum cost. Nevertheless, approximate techniques
provide reasonably good choices and indicate the possibility of
polynomial search cost, on average, for hard random $k$-SAT. The
approximations also explain qualitative features of the algorithm
behavior such as the gradual shift in amplitudes toward low-cost
states and the increasing amplitude variance in the last few steps of
a trial. Moreover, it appears possible to achieve good average
performance with phase parameters depending only on the ensemble
parameters $n$, $k$ and $m$, rather than values tuned to each problem
instance.

The averaging procedure is useful because the quantum algorithm
evaluates the entire search space and hence incorporates information
from all states. By contrast any single run of a classical heuristic
samples only a relatively few states which are unlikely to be typical
of the search space as a whole, hence precluding theoretical analyses
based on average state properties. Thus while quantum heuristics are
difficult to evaluate empirically due to the exponential cost of their
simulation on classical machines, they could allow a simpler, though
still approximate, theoretical analysis than is possible for classical
heuristics.

A number of extensions are possible. First, the amplitude shift of
\fig{shift} means even if a solution is not found after a trial,
the measured state likely has relatively low cost.  Thus, like
local classical search methods such as GSAT but unlike amplitude
amplification, the algorithm applies directly to combinatorial
optimization, i.e., finding a minimal conflict
state~\cite{freuder92}. For example, the shift in amplitudes toward
low-cost states is seen in satisfiability problems with no solutions
and the traveling-salesman problem~\cite{hogg00b}.

Second, the mean-field analysis also applies to other classes of
search problems, provided the probabilities relating problem
properties can be determined. This is possible for a variety of
commonly studied search ensembles such as coloring random
graphs. Ensembles of real-world problems lack analytically known
probability distributions, but sampling representative instances
allows estimating $P(c|c',d)$.  Such estimates may even be useful for
analytically simple ensembles, allowing some tuning of phase
parameters for a particular problem instance. Conversely, which problem classes have so little correlation among
search state properties that quantum algorithms are unlikely to be
particularly useful, on average? Such classes may be useful
for cryptographic applications~\cite{koblitz98}.

Analysis of problem structure can also indicate how the cost varies
through the search space in giving local minima, plateaus,
etc.~\cite{hogg96d,frank97}. Such information may help evaluate other
types of quantum algorithms that rely on properties of the cost
function throughout the space, such as those using partial assignments~\cite{cerf98,hogg97}. 
As another example, a continuous evolution
approach~\cite{farhi00} depends on the nature of the eigenvalue
spectrum of Hamiltonians encoding the problem costs and hence may
benefit from an ensemble analysis of problem structure.

Third, in common with amplitude amplification~\cite{boyer96} and some
classical methods~\cite{luby93a}, the growth of $p^{(h)}(0)$, as seen
in \fig{shift}, means stopping a bit before the largest probability
reduces the expected cost. More generally, a mixture or ``portfolio'' of trials with somewhat different parameter values could give improved trade-offs between expected costs and the variation in costs seen among different instances~\cite{huberman97,gomes97}.

Fourth, the heuristic can readily incorporate other computationally-efficient
properties of the search states as additional arguments to the phase function $\rho$. One such a property is how the number of conflicts in a state compares to those of its neighbors, which is used by a number of conventional heuristics including GSAT.
Moreover, in analogy with quadratically improving conventional heuristics with amplitude amplification~\cite{brassard98}, we could also evaluate a conventional heuristic, such as GSAT, for a fixed number of steps and use the cost of the resulting state to adjust phases (either instead of or in addition to the cost of the original state). In this case we would be searching not for a solution state directly but rather for a ``good'' initial state, i.e., one from which the conventional heuristic rapidly finds a solution. In fact, using just a few steps of GSAT with random SAT instances with $n=12$ and 20 shows the same shift toward low-cost states as seen in \fig{shift}, and the resulting $\Psoln$ is larger. However, for these problem sizes, the $\Psoln$ values in the original algorithm are sufficiently large that even if using a few steps of GSAT were able to increase $\Psoln$ to equal 1, it would not reduce the overall trial cost due to the additional steps involved in evaluating GSAT. Nevertheless, this approach may be useful for larger problem sizes and illustrates the potential trade-off between the cost of the procedure evaluating search state properties and the resulting probability for a solution, which determines the expected number of trials. 
In summary, introducing additional properties in the phase adjustment may give better performance, but increases the possible number of distinct parameter values. Thus numerical optimization of parameter values is likely to be more difficult.

An interesting open question is whether this heuristic can benefit
from using different parameters and numbers of steps for each trial,
as used for amplitude amplification when the number of solutions is
not known. As with amplitude amplification, the simulations indicate a
wide range of performance among different instances with the same $n$
and $m$, even if they have the same number of solutions. This approach
would rely on the variation among problem instances, not addressed by
ensemble averages. Furthermore, the series of low-cost states returned
by the unsuccessful trial may also be useful indications of problem
structure. This provides another contrast with amplitude amplification
where unsuccessful trials simply return randomly selected nonsolution
states, with no bias toward lower costs.

While this discussion is encouraging, we should note its
limitations. The theory does not provide rigorous bounds on the
average search cost. Moreover, even if the algorithm performs well on
average, it has no guarantee for specific instances.  Nevertheless,
restricting consideration to algorithms whose behavior is analytically
simple underestimates the potential of quantum computers for typical
searches, just as is the case for conventional search algorithms.
With ongoing developments in error correction~\cite{shor95,knill98}
and
implementation~\cite{chuang98,chuang98a,kane98,platzman99,mooij99,bayer01,kielpinski01},
quantum machines with even a modest number of bits and limited
coherence time could help address these issues by evaluating
heuristics beyond the range of classical simulation. This will be
particularly useful for more complicated heuristics, using additional
problem properties, whose theoretical analysis is likely to be more
difficult.  Exploring their behavior will identify opportunities
quantum computers have for using information available in
combinatorial searches to significantly improve performance.

\section*{Acknowledgments}
I thank Scott Kirkpatrick for providing the data on the scaling of the
fraction of soluble random 3-SAT problems above the transition point
presented in Ref.~\cite{selman95}.

%%%%%%%%%%%%%%%%%%%%%%%%%%%%%%%%%%%%%%%%%%%%%%%%%%%%%%%%%%%%%%%%%%%%%%
\newpage
\appendix

\section{Derivation of Behavior of $S$}

This appendix describes the derivation of the equation for the
behavior of the spread.

\subsection{Problem Structure}

\begin{figure}[t]
\begin{center}
\epsfbox{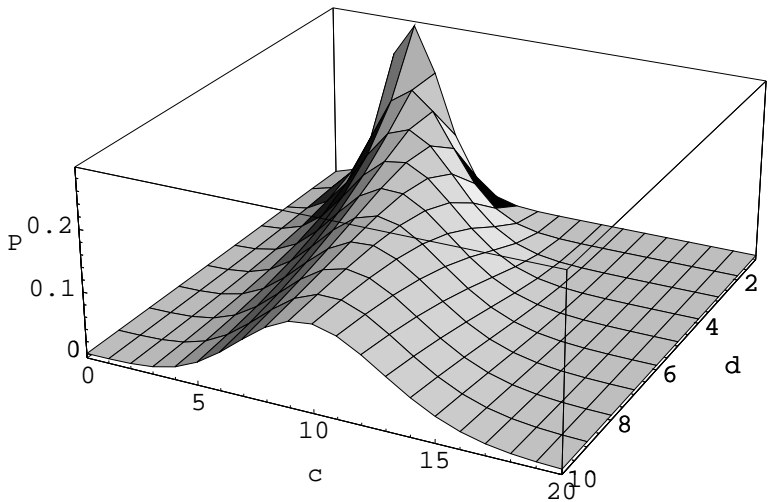}
\end{center}
\caption{\small\label{prob2}Behavior of $P(c|C,d)$ for $n=20$, $k=3$, $m=80$ and $C=3$. The values for $d=0$ are not included: $P(c|C,0)$ is one when $c=C$ and zero otherwise.}
\end{figure}

The algorithm adjusts phases based on the cost associated with each
state, and mixes amplitudes based on Hamming distance between pairs of
states. Evaluating \eq{S mf} requires the relation between distance
and difference in cost. For random $k$-SAT, the required probability
distributions are based on multinomial distributions, which are
approximately gaussian for large problems.

The probability an assignment has cost $C$ is $P(C)={m
\choose C} p^{C} (1-p)^{m-C}$ where $p=2^{-k}$ is the probability a 
single clause conflicts with a given assignment. The expected number
of states with cost $C$ is $v(C) = 2^n P(C)$. As one application, if
the amplitudes after step $h$ satisfy $|\psi_s|^2 \propto a^{c(s)}$ for some constant
$a$, then the probability to obtain a state with $c$ conflicts $p^{(h)}(c)$
is proportional to $P(c) a^c$ giving
\begin{equation}\eqlabel{prob soln}
p^{(h)}(c) = \frac{P(c) a^c}{(1 - p(1-a))^m}
\end{equation}
In particular, $p^{(h)}(0)$ is the probability to obtain a solution.

Similarly, the probability two states separated by distance $d$ have
costs $C$ and $c$, respectively, is given by a sum of multinomials
depending on the number of clauses conflicting with both
states~\cite{hogg00}. The corresponding conditional probability
$P(c|C,d)$ is peaked for $c$ values close to $C$ when $d \ll n$, as
illustrated in \fig{prob2}. As $n$ increases, the relative width of
the probability distribution decreases as $1/\sqrt{n}$, leading to a
high correlation between cost and distance for nearby states. The
expected number of states with $c$ conflicts at distance $d$ from a
state with $C$ conflicts, $\states{d}{C,c}$, is ${n \choose d}
P(c|C,d)$.

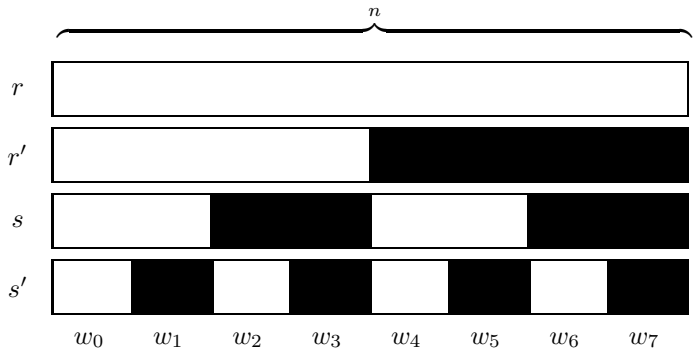
\begin{figure}
\begin{center}
\begin{picture}(260,130)
	\put(145,130){\makebox(0,0){$ \overbrace{\makebox(240,0){}}^n $}}

	\put(20,95){
	\begin{picture}(240,20)
		\put(0,0){\framebox(240,20){}}
	\end{picture}
	}

	\put(20,70){
	\begin{picture}(240,20)
		\put(0,0){\framebox(120,20){}}
		\put(120,0){\framebox(120,20){\rule{120pt}{20pt}}}
	\end{picture}
	}

	\put(20,45) {
	\begin{picture}(240,20)
		\put(0,0){\framebox(60,20){}}
		\put(60,0){\framebox(60,20){\rule{60pt}{20pt}}}
		\put(120,0){\framebox(60,20){}}
		\put(180,0){\framebox(60,20){\rule{60pt}{20pt}}}
	\end{picture}
	}

	\put(20,20) {
	\begin{picture}(240,20)
		\put(0,0){\framebox(30,20){}}
		\put(30,0){\framebox(30,20){\rule{30pt}{20pt}}}
		\put(60,0){\framebox(30,20){}}
		\put(90,0){\framebox(30,20){\rule{30pt}{20pt}}}
		\put(120,0){\framebox(30,20){}}
		\put(150,0){\framebox(30,20){\rule{30pt}{20pt}}}
		\put(180,0){\framebox(30,20){}}
		\put(210,0){\framebox(30,20){\rule{30pt}{20pt}}}
	\end{picture}
	}

	\put(37,10){\makebox(0,0){$w_0$}}
	\put(67,10){\makebox(0,0){$w_1$}}
	\put(97,10){\makebox(0,0){$w_2$}}
	\put(127,10){\makebox(0,0){$w_3$}}
	\put(157,10){\makebox(0,0){$w_4$}}
	\put(187,10){\makebox(0,0){$w_5$}}
	\put(217,10){\makebox(0,0){$w_6$}}
	\put(247,10){\makebox(0,0){$w_7$}}

	\put(10,105){\makebox(0,0){$r$}}
	\put(10,80){\makebox(0,0){$r'$}}
	\put(10,55){\makebox(0,0){$s$}}
	\put(10,30){\makebox(0,0){$s'$}}
\end{picture}
\end{center}
\caption{\small\label{variable groups}
Grouping of variables based on assigned values in assignments $r$,
$r'$, $s$ and $s'$, each shown as a horizontal box schematically
indicating values assigned to each of the $n$ variables. In each
assignment, the value given in $r$ to a variable is shown as white and
the opposite value as black. In this diagram, variables are grouped
according to the differences in values they are given in the four
assignments. For instance, the first group, consisting of $w_0$
variables, has those variables assigned the same value in all four
assignments. The fourth group, with $w_3$ variables, has those
variables with the same values in $r$ and $r'$, but opposite values in
$s$ and $s'$.}
\end{figure}

The quantity $\states{D,d,d',\delta}{C,C',c,c'}$ is the sum, over all
groups of four states $r,r',s,s'$ with the specified distance
relations, of the probability $P(C,C',c,c'|r,r',s,s')$ those states
have, respectively, costs $C,C',c,c'$. For random $k$-SAT, this
probability depends only on the way these states share variables with
the same assigned values, specified by $W=\{w_0,w_1,\ldots,w_7\}$ and
illustrated in \fig{variable groups}. For example, $w_0$ counts the
number of variables assigned the same value in all four states. These
possibilities completely specify the distances between the states,
namely,
\begin{eqnarray}\eqlabel{distances}
D=d(r,r')	&=&	w_4+w_5+w_6+w_7 \\
d=d(r,s)	&=&	w_2+w_3+w_6+w_7 \nonumber \\ 	
d'=d(r',s')	&=&	w_1+w_3+w_4+w_6 \nonumber \\
\delta=d(s,s')	&=&	w_1+w_2+w_5+w_6 \nonumber
\end{eqnarray}
For a given set of values $W$, there are $N(W) = 2^n {n \choose w_0,
\ldots, w_7}$ corresponding choices for the four states.
% D=w_4+w_5+w_6+w_7, other distances: $d=w_2+w_3+w_6+w_7$, $d'=w_1+w_3+w_4+w_6$, $\delta = w_1+w_2+w_5+w_6$, 

Generalizing the case for two states, the probability $P(C,C',c,c'|W)$
is a multinomial sum over the ways the clauses can be selected to
conflict with different subsets of the states, constrained to give the
specified number of conflicts to each of the states. These clause
selections are determined by the principle of inclusion and
exclusion~\cite{palmer85}. Finally,
$\states{D,d,d',\delta}{C,C',c,c'}$ is the sum of $N(W)
P(C,C',c,c'|W)$ over those choices of $W$ matching the specified
distances between the states.

Because of the constraints on the conflicts and distances, the
resulting 4-state probability does not have a simple closed form. It
is nevertheless readily calculated and, for large problems, is
approximately a normal distribution. For use in the expansion
described in the next section, this distribution is multiplied by
powers and summed, which can be done directly using the multinomial
theorem.

\subsection{Expansion for Large Problem Sizes}

\eq{S mf} simplifies in the limit of large $n$ using the following 
observations of its structure. First, for weak mixing, i.e., when
$\tau_h$ is taken to be $\AtMost{1/n}$, the $u_d$ values in \eq{S mf}
decrease as $n^{-d}$ so the main contributions are from terms with
$d,d' \ll n$. Second, as illustrated in \fig{prob2}, nearby states
generally have about the same cost so the $c,c'$ sums in \eq{S mf} are
dominated by the values close to $C,C'$, respectively. Finally, from
\fig{states}, small values for $d,d'$ also require $\delta \sim D$.

Thus to evaluate \eq{S mf}, 
expand $S_{\delta,c,c'}$ as $S_{D,C,C'} Y^{\delta-D} X^{c-C} (X^*)^{c'-C'}$
for values of $\delta,c,c'$ close to $D$ and the dominant $C,C'$. We are particularly interested in the behavior for $D$ near zero.

With this expansion, the sum over $c,c'$ in \eq{S mf} is sum of a
multinomial multiplied by powers, which is readily evaluated for given
distance relations $W$ in terms of the fraction of clauses conflicting
with various subsets of the four states.

For the sums over $d,d'$, the restrictions $d,d' \ll n$ mean the
variable groups shown in \fig{states} are all much less than $n$
except possibly for the two groups contributing to neither the value
of $d$ nor $d'$. From \eq{distances} these are $w_0 = n-D-w_1-w_2-w_3$
and $w_5=D-w_4-w_6-w_7$. This observation, combined with the
contributions from the $u_d$ factors in \eq{S mf}, allow the $d,d'$
sums to be approximated as exponentials.

For evaluating the probability in states with cost $C$ at step $h$ we
need only $S^{(h)}_{0,C,C}$ whose value and change from one step to
the next is determined by the behavior for $D \ll n$ and hence $C'$
close to $C$. Furthermore, the bulk of the probability is concentrated
in states with a narrow range of costs. Thus we can focus on the
behavior near the dominant $C$ value at each step.

Let $X = r e^{i \theta}$ with $r$ and $\theta$ real-valued. With a
narrow distribution of costs, the dominant $C$ equals the average,
i.e., $\sum_C C P(C) S(0,C,C) \propto \sum_C C P(C) r^{2C}$. Thus the
dominant $C$ equals~\cite{hogg00} $r^2 \nu m$ where $\nu =
\frac{p}{1-p(1-r^2)}$.  Hard random $k$-SAT problems have $m \propto
n$, so significant amplitude is in low-cost states whenever $r$ is of
order $1/\sqrt{n}$.  When $r \ll 1/\sqrt{n}$, the lowest cost states
(i.e., the solutions if the problem instance is soluble) have most of
the amplitude.

Expanding around the dominant $C$ value then produces \eq{deqn}.

% for submission insert result of bibtex directly into this file
%\bibliography{physrev,received}
%\bibliographystyle{plain}

\end{document}